\documentclass[prd,amsmath,amssymb,superscriptaddress,floatfix,nofootinbib,10pt]{revtex4}
\usepackage{times}
\usepackage{amssymb,amsbsy,amsmath,amsfonts}
\usepackage{graphicx}
\usepackage{float}
\usepackage{color}
\usepackage{orcidlink}
\usepackage{morefloats}
\usepackage{rotating}
\usepackage{srcltx}
\usepackage{slashed}
\usepackage{multirow}
\usepackage{verbatim}
\usepackage{hyperref}
\usepackage{tabularx}
\usepackage{bm}
\allowdisplaybreaks[4]

\usepackage{bbding}
\usepackage{threeparttable}

\usepackage{mathrsfs}

\DeclareUnicodeCharacter{2212}{\ensuremath{-}}

\newcommand{\PreserveBackslash}[1]{\let\temp=\\#1\let\\=\temp}
\newcolumntype{C}[1]{>{\PreserveBackslash\centering}p{#1}}
\newcolumntype{R}[1]{>{\PreserveBackslash\raggedleft}p{#1}}
\newcolumntype{L}[1]{>{\PreserveBackslash\raggedright}p{#1}}

\begin{document}

\title{Final state interaction in the $\Lambda_b\to D^+D^- \Lambda,~D^0D_s^- p,~D_s^+D_s^-\Lambda$ reactions}

\author{Jing Song\,\orcidlink{0000-0003-3789-7504}}
\email[]{Song-Jing@buaa.edu.cn}
\affiliation{School of Physics, Beihang University, Beijing, 102206, China}
\affiliation{Departamento de Física Teórica and IFIC, Centro Mixto Universidad de Valencia-CSIC Institutos de Investigación de Paterna, 46071 Valencia, Spain}

\author{Melahat Bayar\, \orcidlink{0000-0002-5914-0126 }}
\email[]{melahat.bayar@kocaeli.edu.tr}
\affiliation{Department of Physics, Kocaeli Univeristy, 41380, Izmit, Turkey}
\affiliation{Departamento de Física Teórica and IFIC, Centro Mixto Universidad de Valencia-CSIC Institutos de Investigación de Paterna, 46071 Valencia, Spain}

\author{ Eulogio Oset\,\orcidlink{ https://orcid.org/0000-0002-4462-7919}}
\email[]{oset@ific.uv.es}
\affiliation{Departamento de Física Teórica and IFIC, Centro Mixto Universidad de Valencia-CSIC Institutos de Investigación de Paterna, 46071 Valencia, Spain}
\affiliation{Department of Physics, Guangxi Normal University, Guilin 541004, China}

\begin{abstract}
We study the three-body decays \(\Lambda_b \to D^+ D^- \Lambda,~D^0D_s^- p,~D_s^+D_s^-\Lambda\), considering both internal and external emission mechanisms that produce various meson-baryon final states. Using a unitarized coupled-channel approach based on the local hidden gauge formalism, we analyze final state interactions in the exotic anticharm systems with  strangeness  \(S=-1\) and \(S=-2\). Significant threshold enhancements appear in both pseudoscalar and vector meson-baryon channels, indicating strong final state interactions and the possible formation of molecular exotic states below threshold. These features remain stable under variations of model parameters. Our results provide valuable insights into heavy hadron decay dynamics and offer theoretical guidance for future experimental searches of open-charm strange exotic hadrons.

\end{abstract}

\maketitle


\section{Introduction}\label{sec:Intr}

The discovery of exotic meson and baryon states, which challenge the traditional classification of mesons as $q\bar{q}$ and baryons as three-quark ($qqq$) states, has significantly advanced the field of hadron physics, particularly hadron spectroscopy. This development has motivated extensive theoretical and experimental efforts, and numerous review articles have been devoted to this topic~\cite{Chen:2016qju,Lebed:2016hpi,Esposito:2016noz,Guo:2017jvc,Ali:2017jda,Klempt:2007cp,Klempt:2009pi,Brambilla:2010cs,Olsen:2014qna,Oset:2016lyh,Chen:2016spr,Hosaka:2016pey,Dong:2017gaw,Olsen:2017bmm}.

The study of weak decays of heavy baryons offers valuable insights into the internal structure of hadrons and the dynamics of hadron-hadron interactions. In particular, the three-body decay $\Lambda_b \to D^+ D^- \Lambda$ provides an ideal platform to investigate meson-baryon interactions involving open charm mesons and strange baryons. The LHCb Collaboration has recently reported the measurement of this decay~\cite{LHCb:2024hfo}, making available the invariant mass distributions of the $D^+ \Lambda$, $D^- \Lambda$, and $D^+ D^-$ subsystems. These distributions offer rich opportunities to search for exotic hadronic states dynamically generated through final state interactions (FSI). 
Related to the $\Lambda_b \to D^+ D^- \Lambda$ reaction one would also have $\Lambda_b\to D^0D_s^- p,~D_s^+D_s^-\Lambda$.
The final states in these decays contain subsystems with strangeness $S = -1$ and $S = -2$, including $D^- \Lambda$, $D_s^- p$, and $D_s^- \Lambda$. The quark content of the $D^- \Lambda$ system is $\bar{c} d s u d$, which does not contain any $q\bar q$ couple that could annihilate, and hence is a clear exotic state which cannot be accommodated with a 3 $q$ system. A recent theoretical study~\cite{Song:2025tha}, based on the interaction between heavy anticharmed mesons and octet baryons, predicts a bound state   around 2888 MeV. This result is obtained using the local hidden gauge approach with a commonly adopted cutoff range of $q_{\text{max}} = 550$–$650$ MeV~\cite{Oset:1997it,Debastiani:2017ewu,Feijoo:2022rxf}. The predicted state lies about 20 MeV below the $D_s^- p$ threshold  and about 92 MeV below the $D^- \Lambda$ threshold (strangeness $S = -1$, isospin $I = 1/2$). Bound states of these systems, using the same coupled channels, are also obtained in~\cite{Yalikun:2021dpk}, although less bound. A discussion on these differences is given in~\cite{Song:2025tha}. These findings suggest that the $D^- \Lambda$ and $D_s^- p$ interactions may produce threshold enhancements or cusp-like structures in their respective invariant mass distributions, which could signal the formation of near-threshold molecular states.

The theoretical framework employed here is based on the local hidden gauge approach extended to heavy flavors~\cite{Bando:1984ej,Bando:1987br,Meissner:1987ge,Nagahiro:2008cv}. In this model, the interaction between mesons and baryons is driven by vector meson exchange. The amplitudes for the coupled channels are unitarized by solving the Bethe-Salpeter equation~\cite{Oset:1997it,Kaiser:1995eg,Oller:1997ti,Oller:2000fj,Jido:2003cb,Hyodo:2008xr,Hyodo:2011qc,Sekihara:2014kya,Dong:2021bvy,Dong:2021rpi,Dong:2021juy,Guo:2013sya,Garcia-Recio:2003ejq,Wang:2019spc}, allowing for the generation of molecular states. 

Among many others resonances, this framework has successfully explained the nature of   the $D_{s0}^*(2317)$, which is widely considered a $DK$ hadronic molecule~\cite{vanBeveren:2003kd,Barnes:2003dj,Gamermann:2006nm,Guo:2006rp,Guo:2006fu,Yang:2021tvc,Liu:2022dmm}. Its low mass and narrow width, inconsistent with quark model expectations~\cite{Godfrey:2003kg,Colangelo:2003vg}, are naturally described within this approach. 
Furthermore, the formation of hadronic molecules in weak decays has been observed or suggested in several reactions~\cite{Sakai:2017hpg,Liang:2014ama} (see review paper~\cite{Oset:2016lyh}).

In the $\Lambda_b \to D^+ D^- \Lambda$ decay, the weak transition $b \to c \bar{c} s$ can proceed through different mechanisms, including internal and external emission~\cite{Chau:1982da}. The hadronization of the resulting quark pairs produces various meson-baryon combinations. Internal emission processes often generate meson-meson pairs, while external emission tends to favor meson-baryon configurations. Both types of decay topologies are considered in our analysis, as they lead to different initial meson-baryon combinations and influence the strength of FSI.

The primary objective of this work is to analyze the mass distributions of the $D^- \Lambda$, $D_s^- p$, and $D_s^- \Lambda$ systems in the $\Lambda_b $ decay and search for signs of near-threshold exotic structures. The results are compared with phase space distributions to highlight possible enhancements caused by strong interactions.  These findings may serve as useful input for future experimental measurements. Besides the pseudoscalar meson-baryon systems, this work also explores vector meson-baryon interactions, including $D^{*-} \Lambda$, $D_s^{*-} p$, and $D_s^{*-} \Lambda$. These vector channels, show threshold effects that complement those found in pseudoscalar cases and contribute additional insight into the formation of exotic molecular states. Considering both pseudoscalar and vector mesons allows for a more comprehensive understanding of the heavy meson-baryon dynamics and the possible spectrum of exotic hadrons.

In summary, we present a comprehensive study of the $\Lambda_b \to D^+ D^- \Lambda,~D^0D_s^- p,~D_s^+D_s^-\Lambda$ decays, focusing on the role of final state interactions and their potential to generate exotic hadronic molecules. By incorporating both internal and external emission mechanisms and analyzing all relevant meson-baryon combinations using a unitarized coupled-channel approach, we aim at shedding light on the structure of possible near-threshold states and the dynamics of heavy baryon decays.

\section{Formalism}

In this section, we present the theoretical framework to study the \(\Lambda_b \) decay, emphasizing the role of FSI that can dynamically generate hadronic molecular states. The weak decay of $\Lambda_b$ is initiated by the weak transition of the \(b\) quark   via external and internal $W^-$ emission mechanisms, both of which allow the formation of meson-baryon pairs in the final state  through hadronization. These pairs, through rescattering, can undergo strong interaction leading finally to the $D^-\Lambda$, $D_s^-p$ and $D_s^-\Lambda$ states.   

The strong interaction between the final meson-baryon pairs is treated within a coupled-channel unitary approach, allowing for the dynamical generation of hadronic molecular states. Such interactions can lead to enhancements or resonance-like structures in the different invariant mass distributions. In the following, we detail the weak decay processes, identify the relevant hadronic channels, and discuss how the final state rescattering leads to dynamically generated resonances.

\subsection{Weak decay mechanisms and hadronization: external and internal $W^-$ emissions}

\begin{figure}[H]
    \centering
    \includegraphics[width=0.4\textwidth]{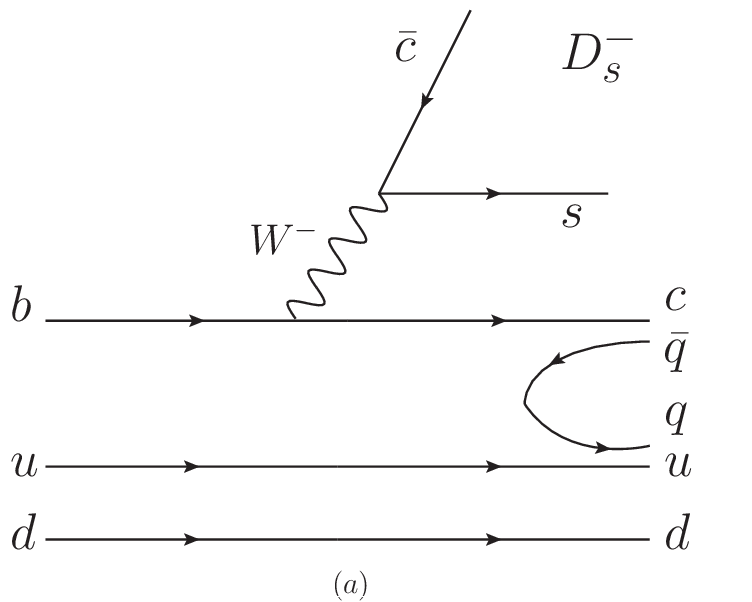}\qquad\qquad
    \includegraphics[width=0.45\textwidth]{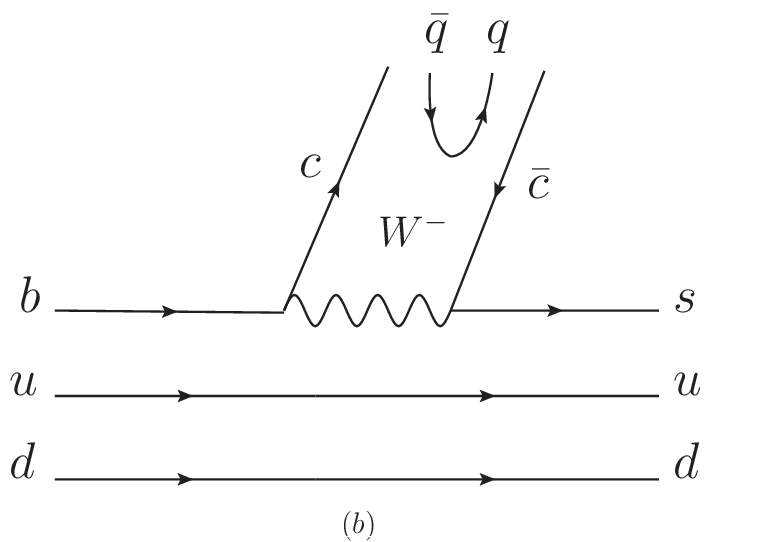}
    \caption{The quark pairs generated through both internal and external $W^-$ emission undergo hadronization to form the final hadrons: (a) external emission,  (b) internal emission.}
    \label{fig:extW}
\end{figure}

In Fig.~\ref{fig:extW} we show the mechanisms of external and internal emission, which after hadronization will produce two mesons and one baryon for the processes that we are interested in.
In the internal $W^-$ emission mechanism Fig.~\ref{fig:extW} (b), the $b$ quark undergos a weak transition via the exchange of a $W^-$ boson producing $c \bar c s$ quarks. The resulting $c \bar c $ quark configuration hadronizes into two mesons.  Although this  Cabibbo-allowed process is color-suppressed, it can play an important role in shaping the decay amplitude through interference and final-state interaction effects.

On the other hand, the external emission mechanism, Fig.~\ref{fig:extW} (a),  proceeds with the $b$ quark decaying weakly into a $c$ quark and a $W^-$ boson, which subsequently produces a $\bar{c}s$ pair, forming a $D_s^-$ meson. 
The remaining $c(ud-du)$ quarks undergo hadonization, involving the $c$ quark, thus producing a $D,~D_s$   meson and three extra quarks which combine to some physical baryon.
This process is Cabibbo-allowed  and color-favored, and generally dominates the overall decay rate.

Let us start from the mechanism of Fig.~\ref{fig:extW} (a) for external emission. We produce directly a $D_s^-$ and we hadronize the $cu$ quark to produce an extra meson baryon state. We have
\begin{align}
\Lambda_b = \frac{1}{\sqrt{2}}b(ud - du)\chi_{MA},
\end{align}
where $\chi_{MA}$ is the mixed antisymmetric spin wave function on the last two quarks. Through weak decay the $\Lambda_b$ becomes 
\begin{align}
\Lambda_b \longrightarrow D_s^-\frac{1}{\sqrt{2}}c(ud - du)\chi_{MA}.
\end{align}
Next we hadronize the $c$ quark and one of the light quarks as 
\begin{align}
\frac{1}{\sqrt{2}}c(ud - du)\longrightarrow \sum_{i} \frac{1}{\sqrt{2}} c \bar q_iq_i(ud-du)\longrightarrow \sum_{i} \frac{1}{\sqrt{2}}  P_{4i}q_i(ud-du),
\end{align}
where $P$ is the $q_i\bar q_j$ matrix written in terms of pseudoscalar mesons
\begin{equation}\label{PMat}
P =
\begin{pmatrix}
\frac{1}{\sqrt{2}} \pi^0 + \frac{1}{\sqrt{3}} \eta + \frac{1}{\sqrt{6}} \eta' & \pi^+ & K^+ & \bar{D}^0 \\
\pi^- & -\frac{1}{\sqrt{2}} \pi^0 + \frac{1}{\sqrt{3}} \eta + \frac{1}{\sqrt{6}} \eta' & K^0 & D^- \\
K^- & \bar{K}^0 & -\frac{1}{\sqrt{3}} \eta + \sqrt{\frac{2}{3}} \eta' & D_s^- \\
D^0 & D^+ & D_s^+ & \eta_c
\end{pmatrix},
\end{equation}
which implies the $\eta,~\eta'$  mixing of Ref.~\cite{Bramon92}.
Ignoring $\eta_c$ which does not play a role here, we have
\begin{align}\label{eq.1h}
\Lambda_b \longrightarrow D_s^-\frac{1}{\sqrt{2}} \bigg\{D^0~u(ud - du)+D^+~d(ud - du)+D_s^+~s(ud - du)\bigg\}\chi_{MA}.
\end{align}
We have to see which baryons come from the contribution of the 3 quarks. Recall that  the baryons of the SU(3) octet are given by
\begin{align}
\phi = \frac{1}{\sqrt{2}}( \phi_{MS}\chi_{MS}+\phi_{MA}\chi_{MA}).
\end{align}
The overlap of $\phi$ with the three quark obtained in Eq.~\ref{eq.1h} is
\begin{equation}
   \Bigg\langle~\phi~ \Bigg| \left\{ 
         \begin{aligned}
           & u \\
           & d \\
           & s \\
         \end{aligned} \right\}(ud - du)~\chi_{MA}~\Bigg\rangle = \frac{1}{\sqrt{2}}
   \Bigg\langle~\phi_{MA}~ \Bigg| \left\{ 
         \begin{aligned}
           & u \\
           & d \\
           & s \\
         \end{aligned} \right\}(ud - du)~~\Bigg\rangle.
\end{equation}
And, consistently with the structure of the Lagrangians  used to describe the meson-baryon interaction, the $\phi_{MA}$  wave functions are given in~\cite{Miyahara:2016yyh}~\footnote{The phase  of the $\Lambda,~\Xi^0,\Sigma^+$ are opposite to those of Ref.~\cite{Close}} as
\begin{align}
&p = \frac{1}{\sqrt{2}}u (ud - du),\\\nonumber
&n = \frac{1}{\sqrt{2}}d (ud - du),\\\nonumber
&\Lambda = \frac{1}{2\sqrt{3}}\left[{u(ds - sd)}+ d(su - us) - {2s(ud - du)}\right], 
\end{align}
Hence, 
\begin{align}\label{eq:ex}
\Lambda_b \longrightarrow \frac{1}{\sqrt{2}}D_s^-D^0p+\frac{1}{\sqrt{2}}D_s^-D^+n-\frac{1}{\sqrt{3}}D_s^-D_s^+\Lambda.
\end{align}
For the internal emission of Fig.\ref{fig:extW} (b) we have the hadronization of the $c\bar c$ pair, which proceeds as 
\begin{align}
c\bar c\longrightarrow \sum_{i} c \bar q_iq_i\bar c&\equiv \sum_{i} P_{4i}P_{i4}=(P^2)_{44}\\\nonumber
&=D^0\bar D^0+D^+D^-+D_s^+D_s^-.
\end{align}
Hence
\begin{align}\label{eq:in}
\Lambda_b &\longrightarrow \frac{1}{\sqrt{2}}(D^0\bar D^0+D^+D^-+D_s^+D_s^-)s(ud-du)\chi_{MA}\\\nonumber
& = - \frac{1}{\sqrt{3}}(D^0\bar D^0+D^+D^-+D_s^+D_s^-)\Lambda.
\end{align}

\begin{figure}[H]
    \centering    \includegraphics[width=0.25\textwidth]{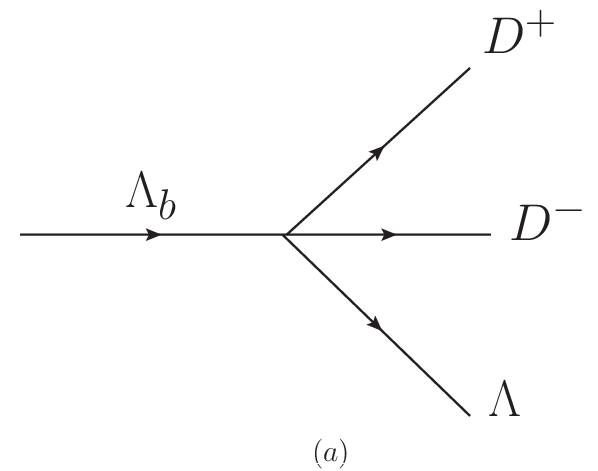}     \includegraphics[width=0.25\textwidth]{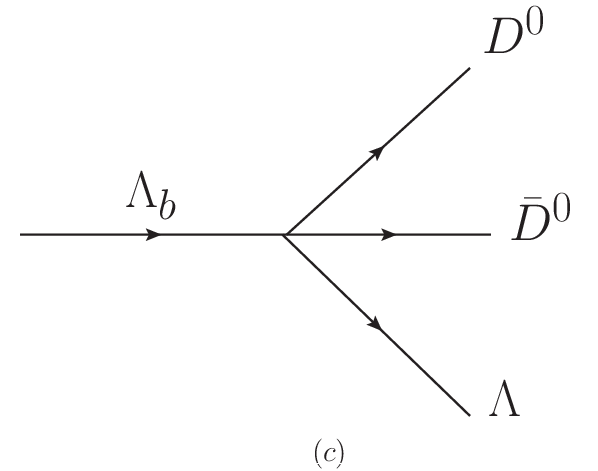}    \includegraphics[width=0.25\textwidth]{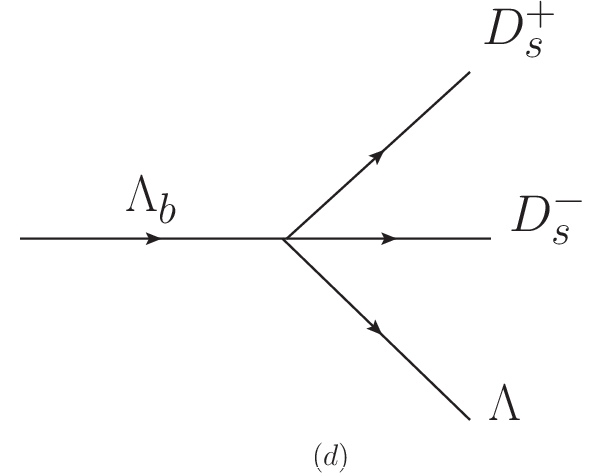}    \includegraphics[width=0.25\textwidth]{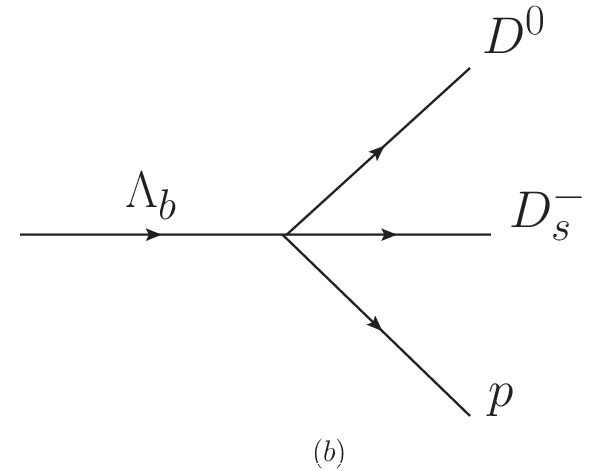}    \includegraphics[width=0.25\textwidth]{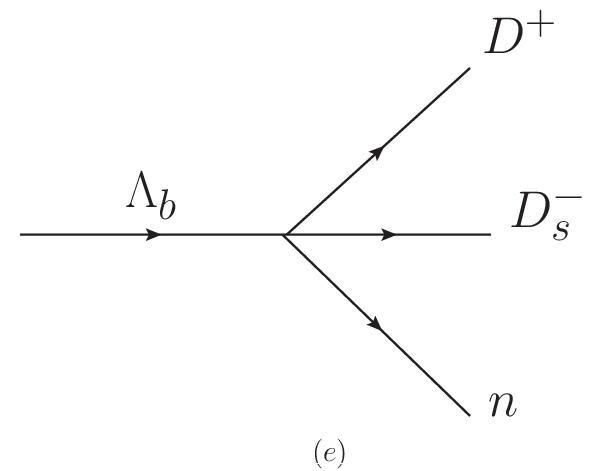}
    \caption{Hadronic configurations from hadronization in the $\Lambda_b $ decay direct production. }
    \label{fig:hadconfig1}
\end{figure}

From both mechanisms, we can expect the production of meson-baryon pairs such as $D^- \Lambda$, $D_s^- p$, and $D_s^- \Lambda$, either directly or through rescattering. These pairs are subject to strong interaction and can form molecular states or cusp-like enhancements. The possible hadronic configurations are shown in Fig.~\ref{fig:hadconfig1}.

\begin{figure}[H]
    \centering
    \includegraphics[width=0.3\textwidth]{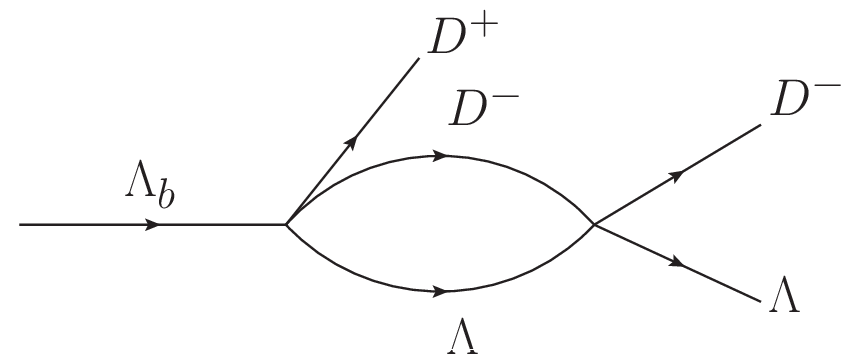}    \includegraphics[width=0.3\textwidth]{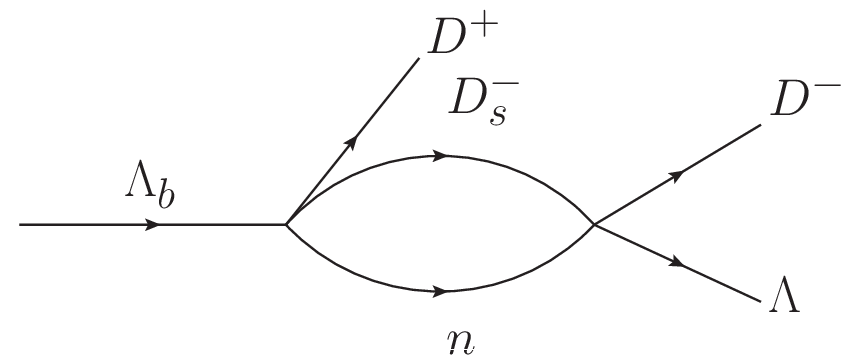}
    \includegraphics[width=0.3\textwidth]{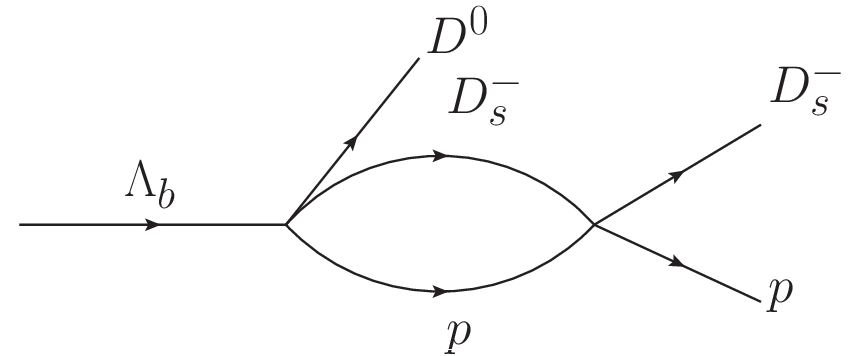}
    \includegraphics[width=0.3\textwidth]{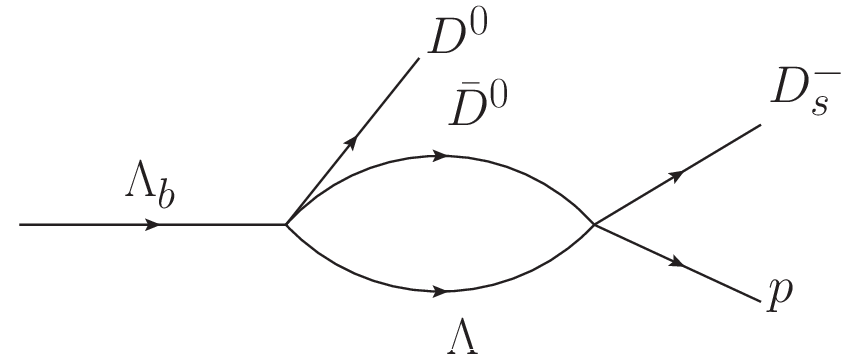}
    \includegraphics[width=0.3\textwidth]{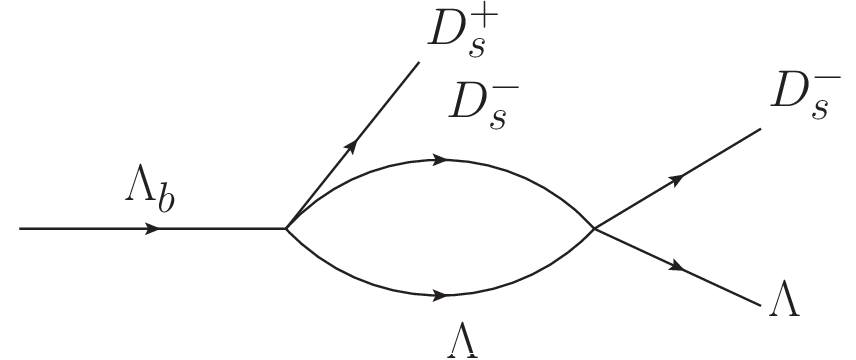}
        \caption{Hadronic configurations from hadronization through rescattering in the $\Lambda_b $ decay. }
    \label{fig:hadconfig2}
\end{figure}

However, we can also obtain the same final states through rescattering, as shown in Fig.~\ref{fig:hadconfig2}.

We next study in detail the interactions among these configurations, starting from the most relevant one: the $D^- \Lambda$ system.

\subsection{The $D^- \Lambda$, $D_s^- p$ coupled channel, and $D_s^- \Lambda$ state in the coupled channels unitary approach}

In the final state, the $D^- \Lambda$ pair can undergo strong interactions, which can be studied using the local hidden gauge approach extended to the heavy quark sector~\cite{Hofmann:2005sw,Yan:2023ttx,Yalikun:2021dpk,Song:2025tha}. 
The $D^- \Lambda$ system couples to the channels $D_s^- p$ and $\bar D\Sigma$ ($I=1/2$).

We consider the  coupled-channel meson-octet baryon interaction  including coupled-channel dynamics corresponding to $|MB\rangle = \{\bar{D}_s N,~ \bar{D} \Lambda,~ \bar{D} \Sigma \}$ and $|MB\rangle = \bar{D}_s \Lambda,~ \bar{D} \Xi \}$ in $S=-1,~I=1/2$ and $S=-2,~I=0$ sectors~\cite{Song:2025tha}, respectively, within the framework of the local hidden gauge approach~\cite{Bando:1984ej,Bando:1987br,Meissner:1987ge,Nagahiro:2008cv}. The $s$-wave interaction kernel is obtained from vector-meson exchange, and the scattering amplitudes are computed by solving the Bethe-Salpeter equation in coupled channels,
\begin{align}
    T = [1 - V G]^{-1} V,
\end{align}
where $V$ is the interaction kernel,
\begin{equation}
V_{ij} = C_{ij} \frac{1}{4f_\pi^2}(k^0 + k^{\prime 0}).
\label{kernel}
\end{equation}
Here, \( k^0 \) and \( k^{\prime 0} \) represent the meson energies in the initial and final states, respectively, and are given by
\begin{equation}
k^0 = \frac{s + m_{m_i}^2 - M_{B_i}^2}{2\sqrt{s}}, \quad 
k^{\prime 0} = \frac{s + m_{m_j}^2 - M_{B_j}^2}{2\sqrt{s}},
\end{equation}
where \( m_{m_i}, M_{B_i} \) and \( m_{m_j}, M_{B_j} \) denote the meson and baryon masses in the initial and final channels, respectively. The interaction strength between channels is characterized by the coefficients \( C_{ij} \), which are listed in Tables~I and II of Ref.~\cite{Song:2025tha}.
The meson-baryon loop function \( G_i(\sqrt{s}) \) is evaluated using a cutoff regularization in momentum space:
\begin{equation}
G_i (\sqrt{s}) = 2M_i \int_{|\vec{q}| < q_{\text{max}}} \frac{d^3 q}{(2\pi)^3} \frac{\omega_1(\vec{q}) + \omega_2(\vec{q})}{2 \omega_1(\vec{q}) \omega_2(\vec{q})} \frac{1}{s - [\omega_1(\vec{q}) + \omega_2(\vec{q})]^2 + i\epsilon},
\label{eq:G}
\end{equation}
where the on-shell energies of the meson and baryon in the loop are given by
\[
\omega_1(\vec{q}) = \sqrt{\vec{q}^{\,2} + m_i^2}, \quad 
\omega_2(\vec{q}) = \sqrt{\vec{q}^{\,2} + M_i^2}.
\]
We adopt a cutoff range of \( q_{\text{max}} = 550-650 \) MeV, in accordance with Ref.~\cite{Oset:1997it,Debastiani:2017ewu,Feijoo:2022rxf}.
 Poles of the $T$ matrix in the complex energy plane correspond to dynamically generated resonances.

\subsection{$D^- \Lambda$ mass distribution in the $\Lambda_b\to D^+D^- \Lambda$ decay}

Within the theoretical framework, the analysis of the $D^- \Lambda$ invariant mass distribution in the decay $\Lambda_b \to D^+ D^- \Lambda$ provides a valuable approach to investigate the dynamics of the $D^- \Lambda$ interaction. In particular, this observable is sensitive to possible resonant structures dynamically generated through final-state interaction. Such structures may manifest as enhancements or peaks near the $D^- \Lambda$ threshold, which can be interpreted as signals of strong interaction effects, including cusp phenomena or the formation of hadronic molecular states. 
Notably, the $D^- \Lambda$ pair originates directly from the internal $W^-$ emission mechanism (see Eq.~(\ref{eq:in})). However, it can also come from external emission and rescattering. 
An enhancement in this region would thus reflect the underlying interaction strength and the  existence of a near-threshold molecular configuration.

\begin{figure}[H]
    \centering
    \includegraphics[width=0.23\textwidth]{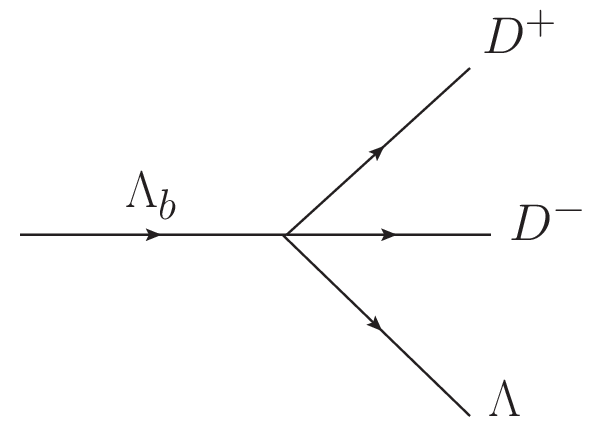}~~
    \includegraphics[width=0.35\textwidth]{Lb_FSI_Feymann_1_1.eps}~~
    \includegraphics[width=0.35\textwidth]{Lb_FSI_Feymann_5_1.eps}
    \caption{Diagrammatic representation of the $D^- \Lambda$ invariant mass distribution in $\Lambda_b $ decay .}
    \label{fig:DmLambda_mass}
\end{figure}

The differential decay width can be written as:
\begin{align}
    \frac{d\Gamma}{dM_\text{inv}({D^- \Lambda})} = \frac{1}{(2\pi)^3} \frac{1}{4 M_{\Lambda_b}^2} p_{D^+}\tilde{p}_{D^-} |{\mathcal M}|^2 \,
\end{align}
with
\begin{align}
&  p_{D^+} = \frac{\lambda^{1 / 2}\left(M_{\Lambda_b}^2, ~m_{D^+}^2,  ~M^2_{\text{inv}}(D^- \Lambda)\right)}{2 M_{\Lambda_b}},\\\nonumber
& \tilde{p}_{D^-} = \frac{\lambda^{1 / 2}\left(M_{\text{inv}}^2(D^- \Lambda), ~m_{D^-}^2, ~m_{\Lambda}^2\right)}{2 M_{\text{inv}}(D^- \Lambda)},
\end{align}
where ${\mathcal M}$ denotes the full decay amplitude, which incorporates both the direct weak production mechanism and the contribution from the final-state interaction,  as illustrated in Fig.~\ref{fig:DmLambda_mass}. 
The $T$ matrix is evaluated using the formalism described in the previous subsection.
\begin{align}\label{eq:tie_1}
 \mathcal M = & -\frac{1}{\sqrt{3}}\mathcal{C}-\frac{1}{\sqrt{3}}\mathcal{C}~G_{D^- \Lambda}(M_\text{inv}(D^- \Lambda))~t_{D^- \Lambda, D^- \Lambda}(M_\text{inv}(D^- \Lambda))\\\nonumber
& +\frac{1}{\sqrt{2}}\mathcal{C}N_c~G_{D_s^-n}(M_\text{inv}(D^- \Lambda))~t_{D_s^-n, D^- \Lambda}(M_\text{inv}(D^- \Lambda)),
 \end{align}
where $\mathcal{C}$ is a normalization constant that includes the common factors of the weak and hadronization amplitudes. The function $G_{D^- \Lambda}(M_\text{inv})$ is the loop function for the intermediate $D^- \Lambda$ channel, and $t_{D^- \Lambda, D^- \Lambda}(M_\text{inv})$ is the corresponding unitarized scattering amplitude evaluated at the invariant mass $M_\text{inv}(D^- \Lambda)$. The second term accounts for the FSI of the $D^- \Lambda$ system. The third term represents the contribution from the coupled channel $D_s^- n \to D^- \Lambda$, where $G_{D_s^- n}$ is the loop function and $t_{D_s^- n, D^- \Lambda}$ is the transition amplitude between the two channels. The factor $N_c=3$ is implemented to take into account the color enhancement of external emission versus internal emission.
Note that while the first two diagrams in Fig.~\ref{fig:DmLambda_mass} involve the weak internal emission mechanism, the third diagram (third term in Eq.~(\ref{eq:tie_1})) involves weak external emission. Hence, this term should have a large weight compared to the first two terms, enhancing the role of the coupled channels interaction.

\subsection{ $D_s^- p$  mass distribution in the $\Lambda_b\to D^0D_s^- p$ decay}

\begin{figure}[H]
    \centering
    \includegraphics[width=0.23\textwidth]{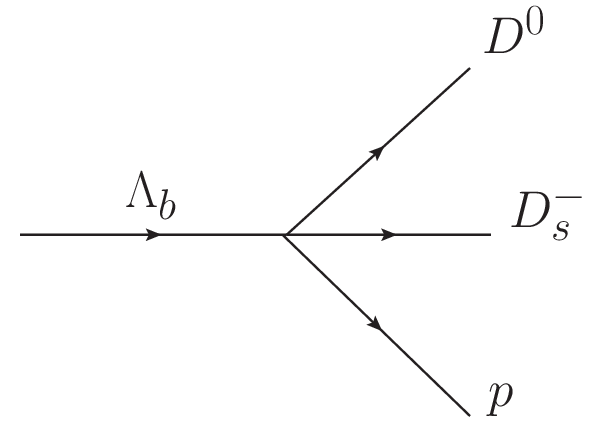}
    \includegraphics[width=0.35\textwidth]{Lb_FSI_Feymann_2_1.eps}
    \includegraphics[width=0.35\textwidth]{Lb_FSI_Feymann_3_1.eps}
    \caption{Diagrammatic representation of the $D_s^- p$ contribution in $\Lambda_b$ decay, including both direct production and rescattering effects.  }
    \label{fig:Dsp_coupling}
\end{figure}

Next we consider a related reaction where the $D_s^- p$ channel is produced in the final state.
As seen in Eq.~(\ref{eq:ex}) the $D^0D_s^- p$ channel appears in the $\Lambda_b$ decay from external emission, but it can also be reached via rescattering from a first step production of $D^0\bar D^0\Lambda$  in internal emission. The mechanisms for  $\Lambda_b\to D^0D_s^- p$  decay are depicted in Fig.~\ref{fig:Dsp_coupling}.

The differential decay width with the $D_s^- p$ channel included can be expressed as
\begin{align}
\frac{d\Gamma}{dM_\text{inv}({D_s^- p})} = \frac{1}{(2\pi)^3} \frac{1}{4 M_{\Lambda_b}^2} p_{ D^0} \tilde{p}_{D_s^-}  |{\mathcal M}|^2,
\end{align}
with
\begin{align}
&  p_{ D^0} = \frac{\lambda^{1 / 2}\left(M_{\Lambda_b}^2, ~m_{ D^0}^2,  ~M^2_{\text{inv}}(D_s^- p)\right)}{2 M_{\Lambda_b}},\\\nonumber
& \tilde{p}_{D_s^-} = \frac{\lambda^{1 / 2}\left(M_{\text{inv}}^2(D_s^- p), ~m_{D_s^-}^2, ~m_{p}^2\right)}{2 M_{\text{inv}}(D_s^- p)},
\end{align}
where the full decay amplitude $\mathcal M$ now contains both the primary weak production amplitude and contributions from the final-state interactions involving the $D_s^- p$ intermediate state. 
Hence, the corresponding amplitudes $ \mathcal M $ is written as,
\begin{align}\label{eq:tie_2}
 \mathcal M = & \frac{1}{\sqrt{2}}\mathcal{C}N_c+\frac{1}{\sqrt{2}}\mathcal{C}N_c~G_{D_s^- p}(M_\text{inv}(D_s^- p))~t_{D_s^- p, D_s^- p}(M_\text{inv}(D_s^- p))\\\nonumber
& -\frac{1}{\sqrt{3}}\mathcal{C} ~G_{\bar D^0\Lambda}(M_\text{inv}(D_s^- p))~t_{\bar D^0 \Lambda, D_s^- p}(M_\text{inv}(D_s^- p)),
 \end{align}
where the first term represents the tree-level contribution to the $D_s^- p$ channel, with $N_c$ being the corresponding color enhancement factor. The second term accounts for the final state interaction of the $D_s^- p$ channel, involving the loop function $G_{D_s^- p}$ and the scattering amplitude $t_{D_s^- p, D_s^- p}$. The third term includes the effect of the coupled-channel transition from $\bar{D}^0 \Lambda$ to $D_s^- p$, with the corresponding loop function $G_{\bar{D}^0 \Lambda}$ and transition amplitude $t_{\bar{D}^0 \Lambda, D_s^- p}$.

\subsection{ $D_s^- \Lambda$  mass distribution in the $\Lambda_b\to D_s^+D_s^- \Lambda$ decay}

\begin{figure}[H]
\centering
\includegraphics[width=0.23\textwidth]{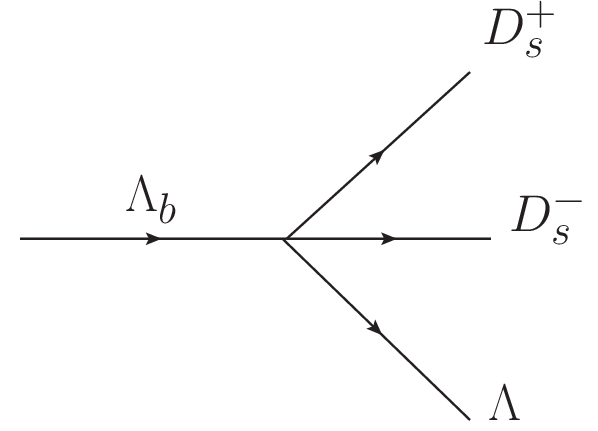}
\includegraphics[width=0.35\textwidth]{Lb_FSI_Feymann_4_1.eps}
\caption{Diagrammatic representation of the $D_s^- \Lambda$ contribution in $\Lambda_b$ decay, including both direct production and rescattering effects.}
\label{fig:DsL_MASSDIS}
\end{figure}

The study of the $D_s^- \Lambda$ invariant mass distribution in the $\Lambda_b\to D_s^+D_s^- \Lambda$ reaction provides an opportunity to explore coupled-channel dynamics in a more exotic sector.
As in previous cases, the effect of final-state interactions is implemented through the corresponding scattering amplitudes. The direct production and  rescattering processes involving the $D_s^- \Lambda$ pair are schematically illustrated in Fig.~\ref{fig:DsL_MASSDIS}.

The differential decay width with respect to the $D_s^- \Lambda$ invariant mass is given by:
\begin{align}
\frac{d\Gamma}{dM_\text{inv}({D_s^- \Lambda})} = \frac{1}{(2\pi)^3} \frac{1}{4 M_{\Lambda_b}^2} p_{D_s^+} \tilde{p}_{D_s^-}  |{\mathcal M}|^2,
\end{align}
with
\begin{align}
&  p_{D_s^+} = \frac{\lambda^{1 / 2}\left(M_{\Lambda_b}^2, ~m_{D_s^+}^2,  ~M^2_{\text{inv}}(D_s^- \Lambda)\right)}{2 M_{\Lambda_b}},\\\nonumber
& \tilde{p}_{D_s^-} = \frac{\lambda^{1 / 2}\left(M_{\text{inv}}^2(D_s^- \Lambda), ~m_{D_s^-}^2, ~m_{\Lambda}^2\right)}{2 M_{\text{inv}}(D_s^- \Lambda)},
\end{align}
where $\mathcal M$  denotes the full decay amplitude, which includes the direct weak decay contribution and the FSI in the $D_s^- \Lambda$ channel. 
\begin{align}\label{eq:tie_2}
 \mathcal M  = -\frac{1}{\sqrt{3}}\mathcal{C}-
 \frac{1}{\sqrt{3}}\mathcal{C} ~G_{D_s^- \Lambda}(M_\text{inv}(D_s^- \Lambda))~t_{D_s^-\Lambda, D_s^- \Lambda}(M_\text{inv}(D_s^- \Lambda)).
 \end{align}
The first term corresponds to the tree-level amplitude, while the second term accounts for the FSI through the loop function $G_{D_s^- \Lambda}$ and the unitarized scattering amplitude $t_{D_s^- \Lambda, D_s^- \Lambda}$, both evaluated at the invariant mass $M_\text{inv}(D_s^- \Lambda)$.

\section{Results} 
\label{res}

In this section, we present the invariant mass distributions of various meson-baryon systems resulting from the three-body decay of $\Lambda_b  $ discussed above. These distributions are a consequence of the near-threshold hadronic molecules and strong FSI that we obtained in our approach. We focus on subsystems such as $D^- \Lambda$, $D_s^- p$, $D_s^- \Lambda$, as well as their vector counterparts $D^{*-} \Lambda$, $D_s^{*-} p$, and $D_s^{*-} \Lambda$. This latter channels were also studied in Ref.~\cite{Song:2025tha} and we use the results from there. The shapes of these mass distributions, especially near threshold, once determined experimentally, can provide valuable information about the nature of the underlying interactions and the possibility of existence of  dynamically generated states. 

To account for theoretical uncertainties, we perform calculations using different values of the cutoff parameter $q_{\mathrm{max}}$ ranging from $550$ to $650\,\mathrm{MeV}$, consistent with earlier works~\cite{Song:2025tha}. For each case, we also compare the results with normalized phase space distributions to highlight dynamical effects beyond simple kinematics. 

In all the calculation the normalization constant is taken as $ \mathcal C =1$. This means that the results are in arbitrary normalization. However, and  this is important, the relative weights of all the mass distributions that we present are a prediction of the theory.

\subsection{Mass Distributions of $D^- \Lambda$ and $D_s^- p$}

\begin{figure}[H]
    \centering
    \includegraphics[width=0.43\textwidth]{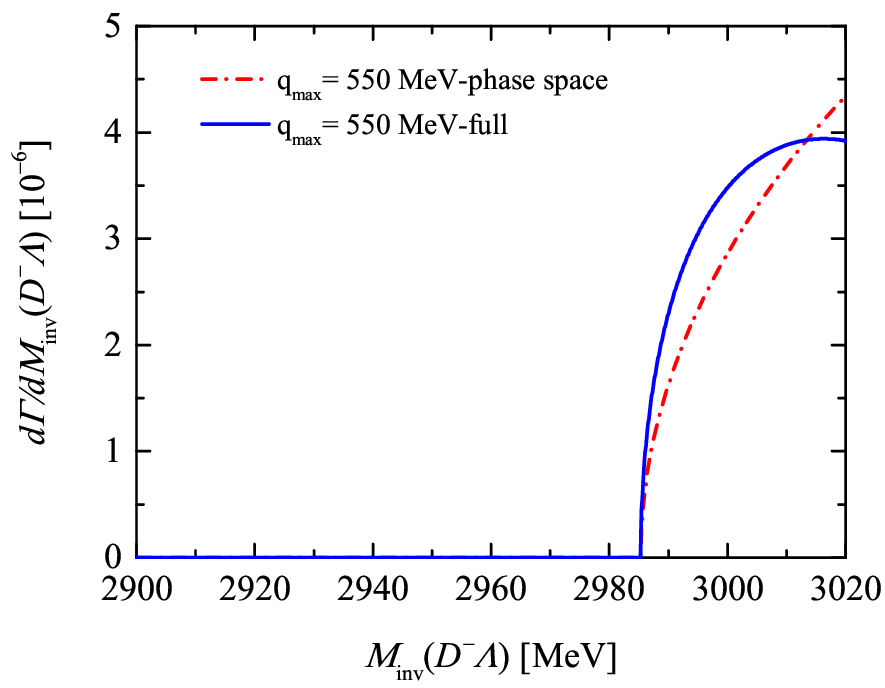}
        \includegraphics[width=0.43\textwidth]{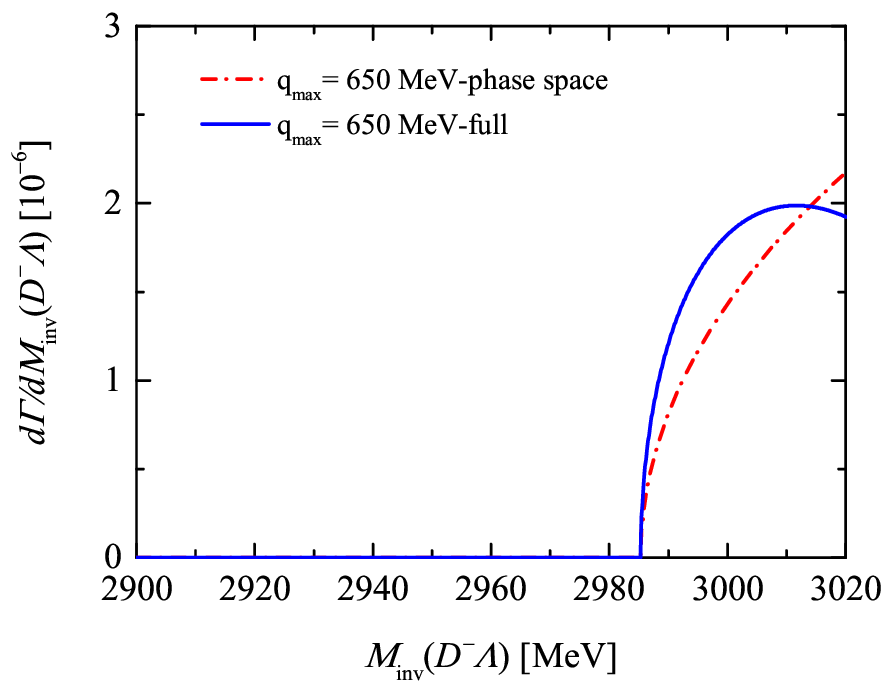}  
\caption{
Invariant mass distributions of \( \frac{d\Gamma}{dM_{\mathrm{inv}}(D^- \Lambda)} \) in the $\Lambda_b \to D^+D^- \Lambda$ decay with \( q_{\text{max}} = 550\,\mathrm{MeV} \) and \( 650\,\mathrm{MeV} \), showing phase space and full contributions. The red  dot-dashed 
  line shows the phase space, while the blue solid  line  represents the full contributions, respectively.}
    \label{fig:DmLambda_DmLambda}
\end{figure}

\begin{figure}[H]
    \centering
    \includegraphics[width=0.45\textwidth]{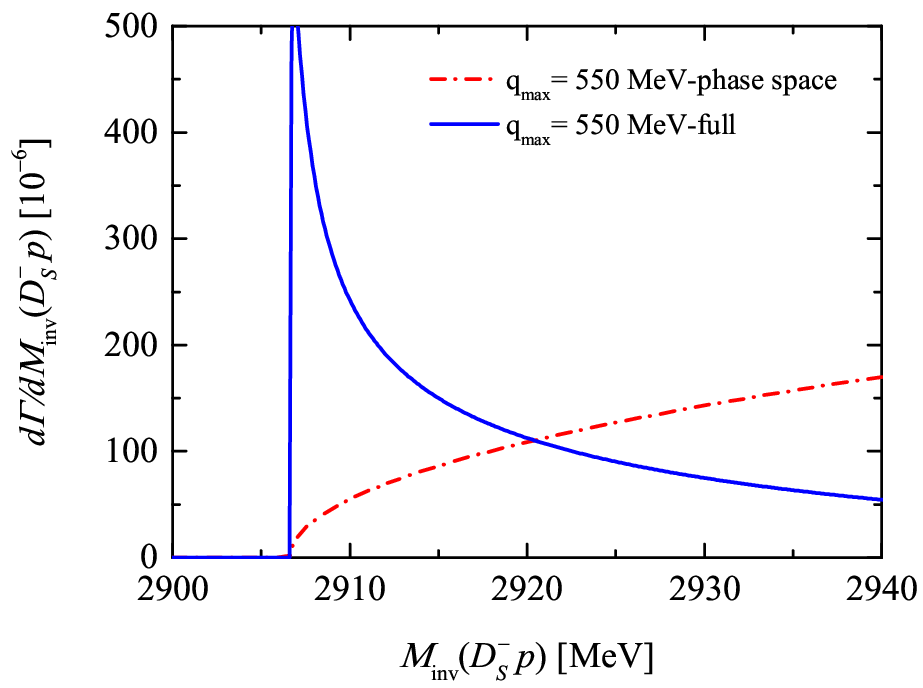}
    \includegraphics[width=0.45\textwidth]{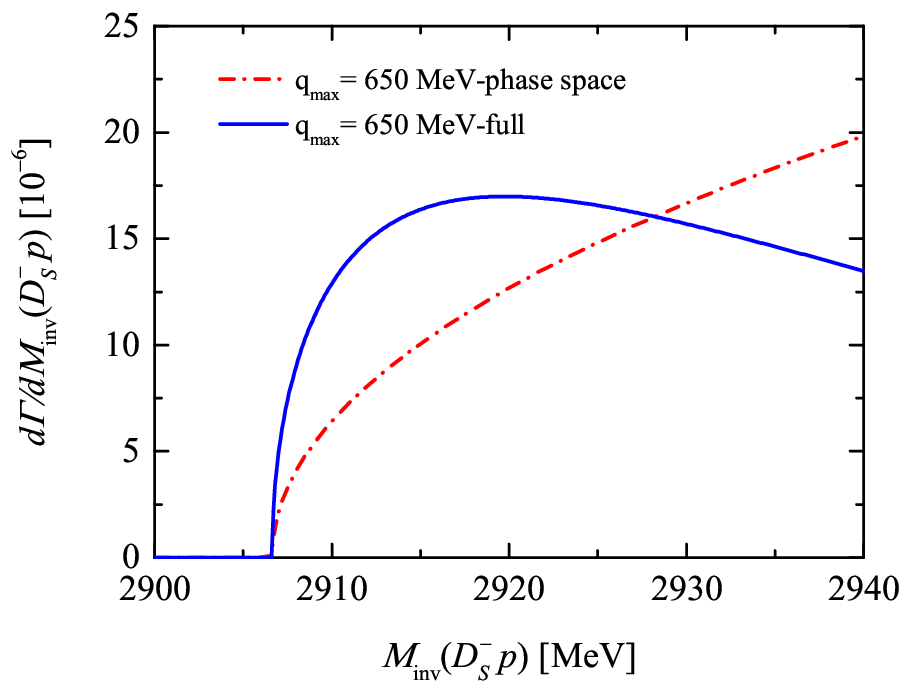}   
\caption{
Invariant mass distribution of \( \frac{d\Gamma}{dM_{\mathrm{inv}}(D_s^- p)} \) in the $\Lambda_b \to D^0D_s^- p$ decay. Line styles and \( q_{\text{max}} \) values are the same as in Figure~\ref{fig:DmLambda_DmLambda}.}
    \label{fig:DmLambda_DsmP}
\end{figure}

Figs.~\ref{fig:DmLambda_DmLambda} and ~\ref{fig:DmLambda_DsmP} show the invariant mass distributions $\frac{d\Gamma}{dM_{\mathrm{inv}}(D^- \Lambda)}$ and $\frac{d\Gamma}{dM_{\mathrm{inv}}(D_s^- p)}$ as functions of their respective invariant masses. We also plot the phase space contribution normalized to the same area in the region chosen of about 40 MeV above threshold. Both distributions for the full mechanism exhibit strong enhancements near threshold, which point to strong final state interactions and the  existence of molecular states of our approach.
To understand these results we show in Table~\ref{tab:results_pole} the results obtained in Ref.~\cite{Song:2025tha} for the  $D^-\Lambda,~D_s^-p$ channel and in $D_s^-\Lambda$.
\begin{table}[H]
\centering
\caption{Poles obtained for $S = -1,~I = 1/2$ and $S = -2,~I = 0$, unit in MeV.} 
\label{tab:results_pole}
\setlength{\tabcolsep}{20pt}
\begin{tabular}{cccc}
\hline \hline
$q_\text{max}$ &  $S = -1,~I = 1/2$ ($D^-\Lambda,~D_s^-p$)   & $q_\text{max}$ &   $S = -2,~I = 0$ ($D_s^-\Lambda$) \\
\hline
550 & 2906 & 560 & 3083\\
650 & 2880 & 650 & 3046\\
\hline
  &  $S = -1,~I = 1/2$ ($D^{*-}\Lambda,~D_s^{*-}p$)   & $q_\text{max}$ &   $S = -2,~I = 0$ ($D_s^{*-}\Lambda$) \\
550 & 3049 & 550 & 3227\\ 
650 & 3020 & 650 & 3184\\
\hline
\hline
\end{tabular}
\end{table}

These results correspond to the case of octet baryons, and consider strangeness $S=-1$ with isospin $I=1/2$. 
From Table~\ref{tab:results_pole}, we can see that for  $q_{\mathrm{max}}=550$ MeV, we find a barely bound state with respect to the $D_s^-p$ threshold (2906.62 MeV). This is the reason for the type of enhancement found in Figs.~\ref{fig:DmLambda_DsmP}, with respect to phase space.  When we use $q_{\mathrm{max}}=650$ MeV, the state is bound by 26 MeV, and the enhancement seen in Fig.~\ref{fig:DmLambda_DsmP} is not so big. Even then there is a clear difference of the shape of the mass distribution with respect to phase space.

Note that with respect to $D^-\Lambda$, the bound states are further away, and hence, as seen in Fig.~\ref{fig:DmLambda_DmLambda}, the enhancement of the mass distribution close to the $D^-\Lambda$  threshold, with respect to phase space is not significative.

Our results show which is the decay mode most favorable to see the predicted bound states of these exotic systems, which is the $\Lambda_b \to D^0D_s^- p$  reaction.
While the exact position and magnitude of the peaks exhibit some sensitivity to the cutoff value, the presence of strong interaction effects near threshold remains well established. These findings encourage further experimental investigation in this invariant mass region, particularly with higher-statistics data.

\subsection{Mass Distribution of $D_s^- \Lambda$}

\begin{figure}[H]
    \centering
\includegraphics[width=0.43\textwidth]{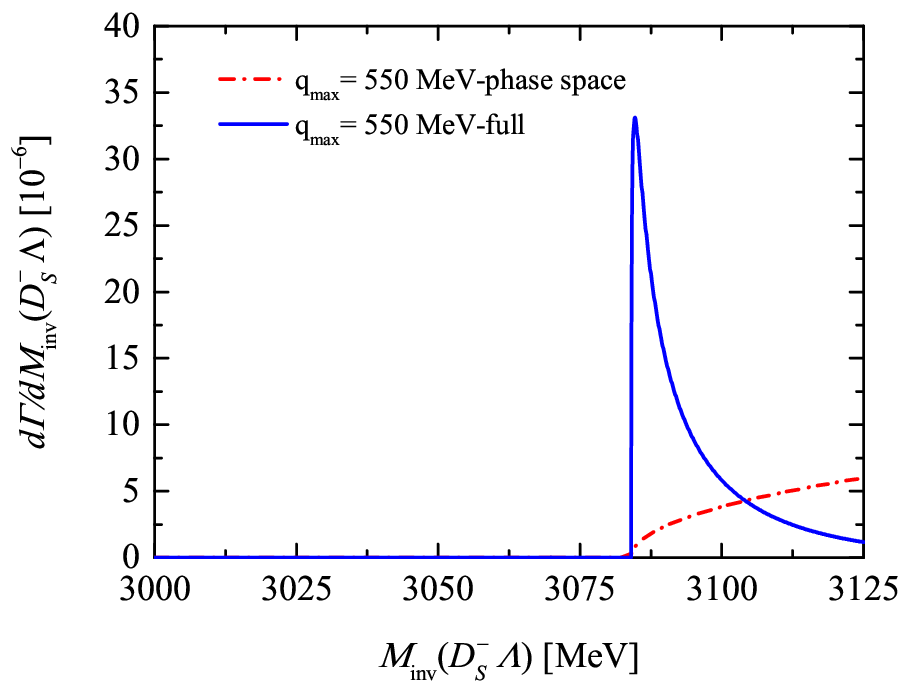}
\includegraphics[width=0.45\textwidth]{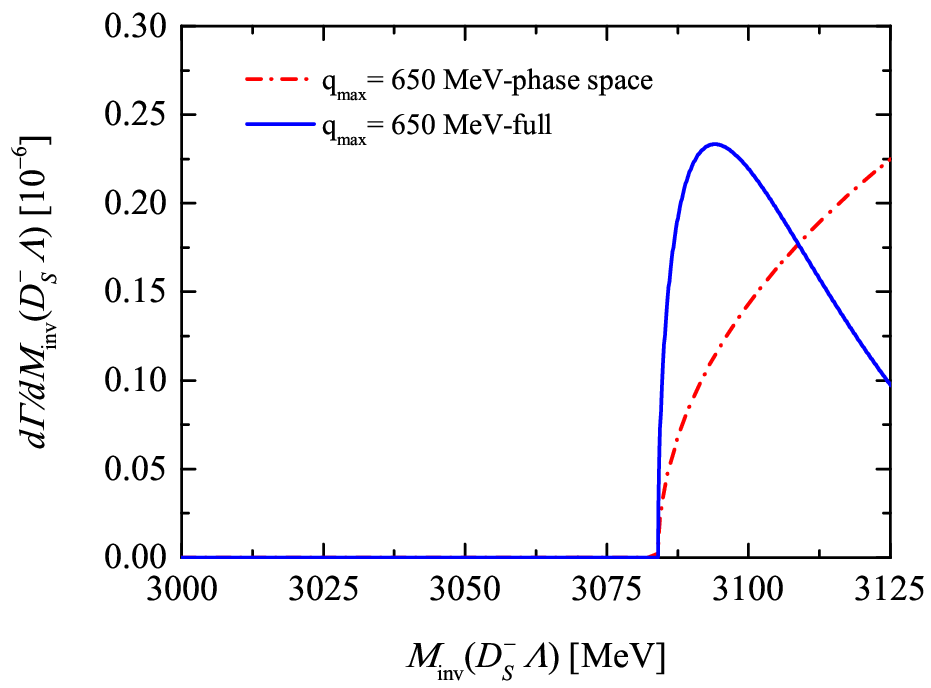}
    
\caption{
Invariant mass distribution of \( \frac{d\Gamma}{dM_{\mathrm{inv}}(D_s^- \Lambda)} \). Line styles and \( q_{\text{max}} \) values are the same as in Figure~\ref{fig:DmLambda_DmLambda}.
}
    \label{fig:DsmLambda}
\end{figure}

Figure~\ref{fig:DsmLambda} shows the invariant mass distribution $\frac{d\Gamma}{dM_{\mathrm{inv}}(D_s^- \Lambda)}$. This channel belongs to the $S = -2$, $I = 0$ sector and provides a unique opportunity to explore doubly-strange hadronic molecules, which have not been extensively studied.
We explore the theoretical uncertainty by varying the cutoff parameter $q_{\mathrm{max}}$ between $550$ and $650\,\mathrm{MeV}$, consistent with previous works using the local hidden gauge approach~\cite{Oset:1997it,Debastiani:2017ewu,Feijoo:2022rxf}.

A   visible  enhancement near threshold is observed for all values of $q_{\mathrm{max}}$, as $q_{\mathrm{max}}$ decreases, the enhancement becomes slightly more pronounced and shifts subtly toward lower invariant masses, similar to the behavior seen in the $S = -1$ channels. 
This is a clear consequence of the pole getting closer to the threshold for smaller value of $q_\text{max}$.

The uncertainties in the position of the poles revert on uncertainties of the mass distributions. Yet, the strong theoretical support for these bound states and the results shown here indicate that the measurements of these mass distributions will be determinant to clarify this issue and find evidence for the existence of these exotic bound states.

\subsection{Mass Distributions of $D^{*-} \Lambda$, $D_s^{*-} p$, and $D_s^{*-} \Lambda$}

\begin{figure}
    \centering
    \includegraphics[width=0.43\textwidth]{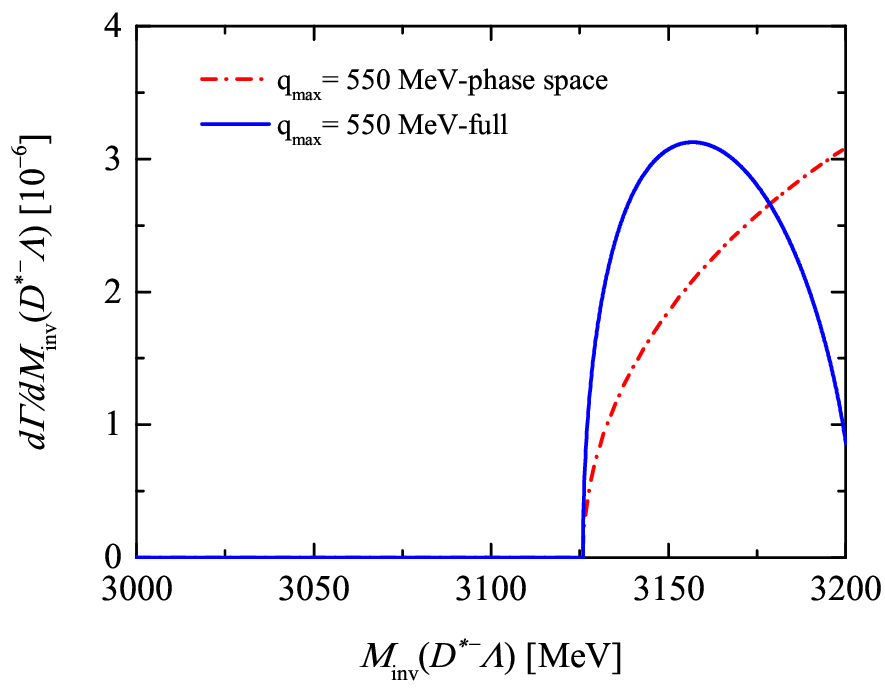}    \includegraphics[width=0.45\textwidth]{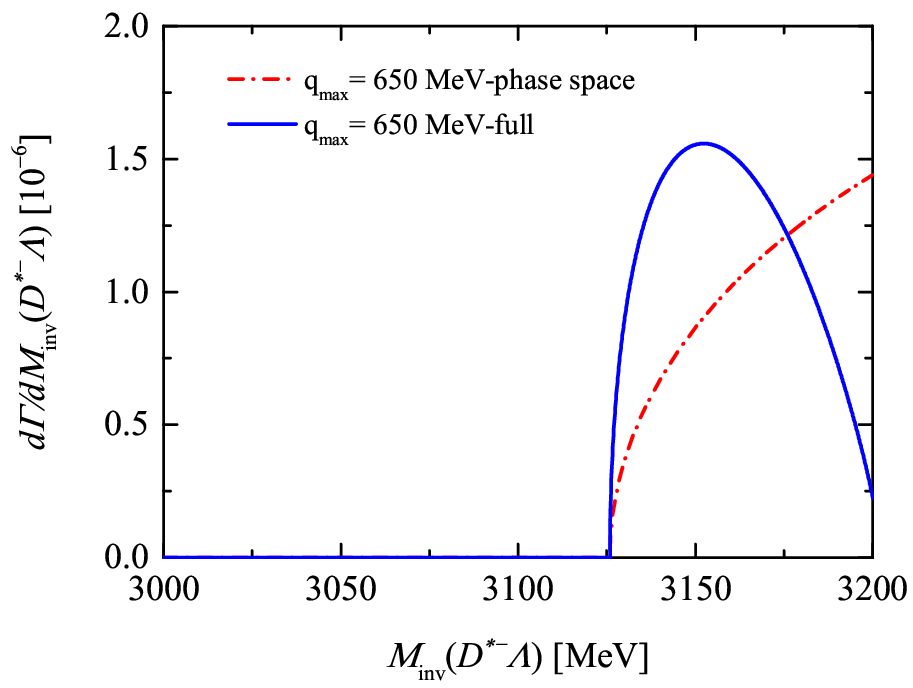}    \includegraphics[width=0.45\textwidth]{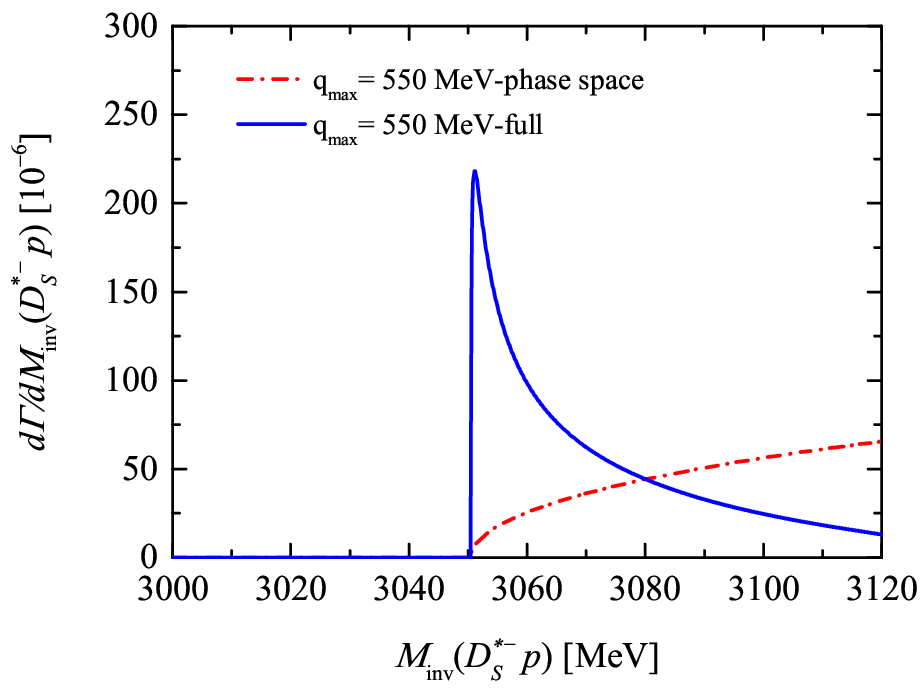}
   \includegraphics[width=0.45\textwidth]{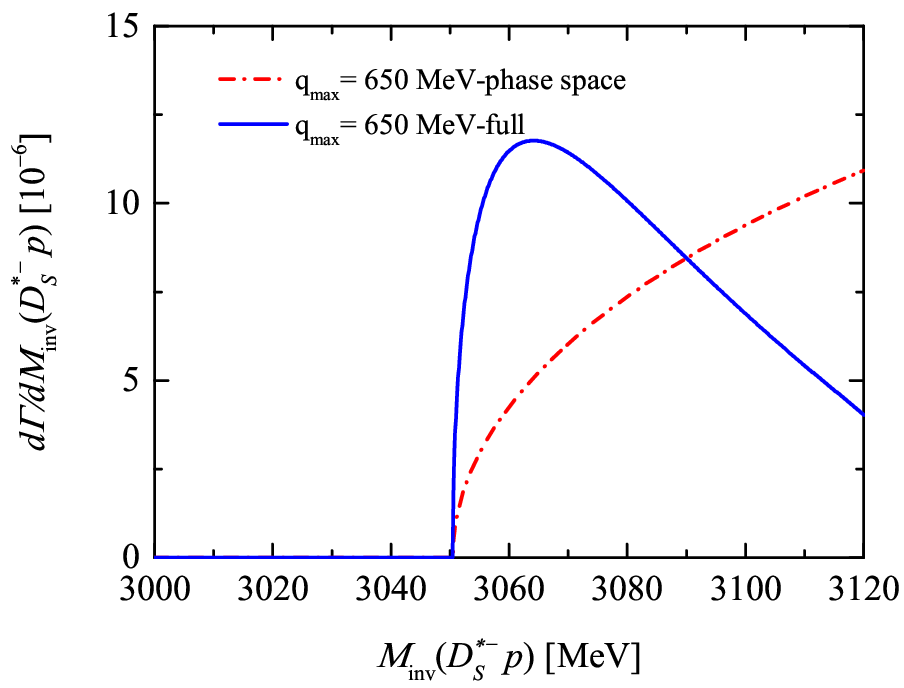}    \includegraphics[width=0.44\textwidth]{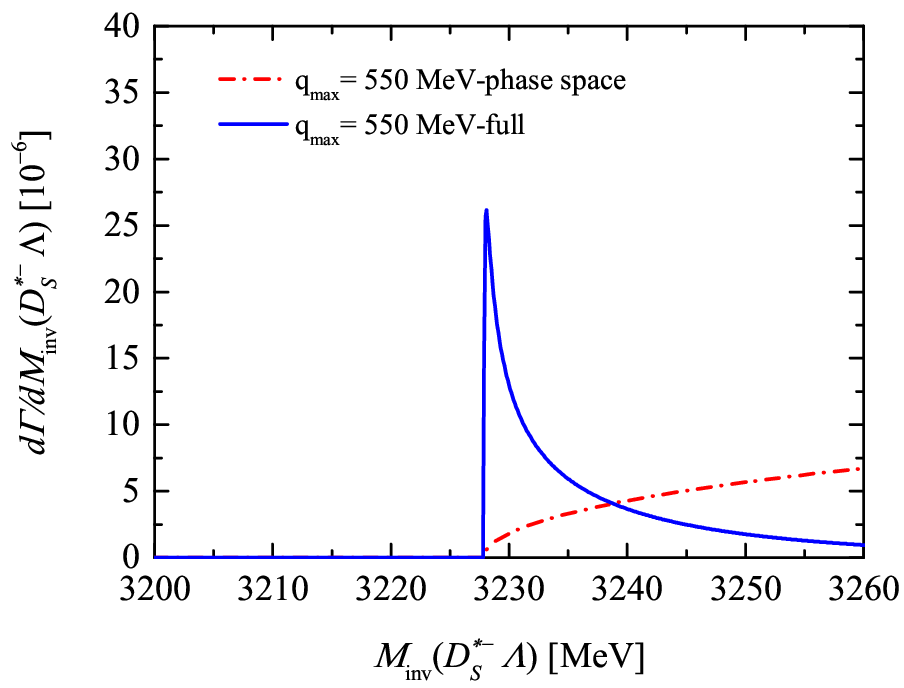}    \includegraphics[width=0.45\textwidth]{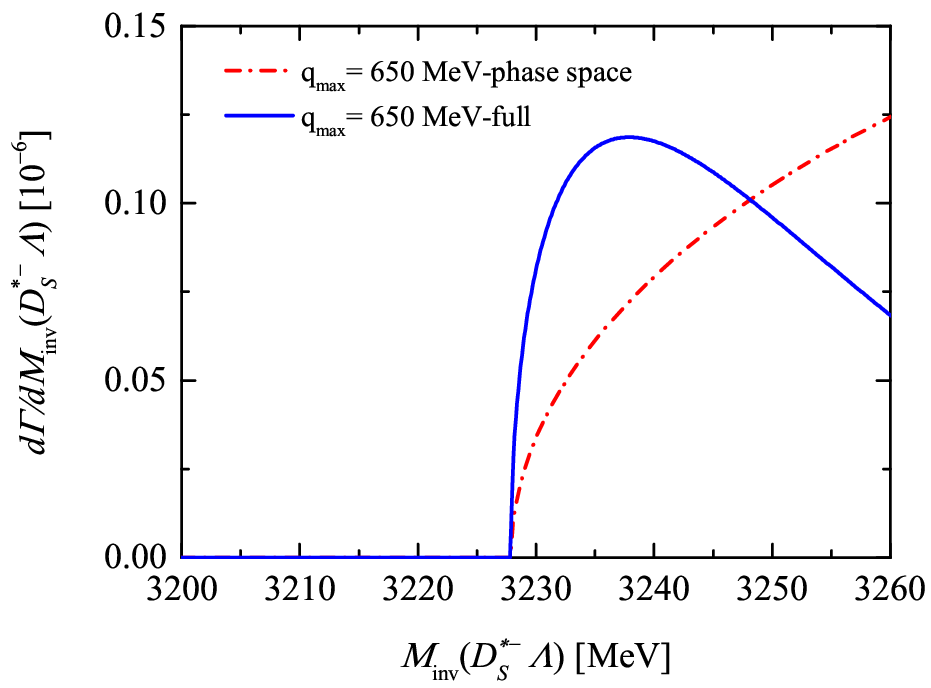}   
\caption{
Invariant mass distributions of \( D^{*-} \Lambda \) in the \( \Lambda_b \to D^{*+} D^{*-} \Lambda \) decay, \( D_s^{*-} p \) in the \( \Lambda_b \to D^{*0} D_s^{*-} p \) decay, and \( D_s^{*-} \Lambda \) in the \( \Lambda_b \to D_s^{*+} D_s^{*-} \Lambda \) decay. Line styles and \( q_{\text{max}} \) values are the same as in Figure~\ref{fig:DmLambda_DmLambda}.}
\label{fig:DstLambda_Dsstp_DsstLambda}
\end{figure}

In Figure~\ref{fig:DstLambda_Dsstp_DsstLambda}, we show the invariant mass distributions $\frac{d\Gamma}{dM_{\mathrm{inv}}}$ for the vector meson-baryon systems: $D^{*-} \Lambda$, $D_s^{*-} p$, and $D_s^{*-} \Lambda$. These channels extend our investigation into vector meson interactions in the same strangeness sectors as the pseudoscalar cases.

The $D^{*-} \Lambda$ and $D_s^{*-} p$ distributions, both belonging to the $S=-1$, $I=1/2$ sector, exhibit noticeable threshold enhancements, as a consequence of the existence of bound states in this sector. The $D_s^{*-} \Lambda$ mass distribution, with $S = -2$, $I = 0$, also shows an  enhancement near threshold.

We again perform an uncertainty analysis by varying \( q_{\mathrm{max}} \) from 550 to 650\,MeV. The corresponding results are displayed in the figures and show that the qualitative features of the threshold behavior are stable.
 The phase space distributions, normalized to the same area, are included for comparison and demonstrate the clear impact of final state interactions on the line shapes.

These findings suggest that vector meson-baryon systems deserve further attention in the search for exotic states, complementing the results found in pseudoscalar channels.

\section{Conclusion} \label{sec:conclusion}

In this work, we have carried out a detailed study of the three-body decay processes \(\Lambda_b \to D^+ D^- \Lambda\), $\Lambda_b \to D^0D_s^- p$, and $\Lambda_b \to D_s^+D_s^-\Lambda$,  focusing on the invariant mass distributions of various meson-baryon subsystems. By employing a unitarized coupled-channel approach based on the local hidden gauge formalism, we have investigated both pseudoscalar and vector meson-baryon interactions in strangeness sectors \(S = -1\) and \(S = -2\), with particular emphasis on the effects of final state interactions on the mass distributions.

Our analysis reveals  enhancements near the thresholds of the \(D^- \Lambda\), \(D_s^- p\), and \(D_s^- \Lambda\) channels, as a consequence of the strong final state interactions and the  existence of molecular exotic states. These threshold effects remain evident despite variations of the cutoff parameter \(q_{\mathrm{max}}\), confirming the stability of the predicted structures.

In addition to pseudoscalar meson-baryon systems, we also investigate vector meson-baryon interactions such as \( \Lambda_b \to D^{*+} D^{*-} \Lambda \), \( D^{*0}D_s^{*-} p \), and \( D_s^{*+}D_s^{*-} \Lambda \). These vector channels, sharing the same strangeness sectors, exhibit threshold behaviors that complement and enrich the search for exotic molecular states. This extended study provides a more complete picture of heavy meson-baryon dynamics and highlights the importance of considering both pseudoscalar and vector mesons in understanding the spectrum of possible exotic hadrons.

The comparison with normalized phase space distributions clearly highlights the dynamical origin of the observed mass distribution features. The study of these decay channels offers valuable insights into the interactions of multi-strange meson-baryon systems and could uncover exotic states beyond the conventional hadron framework.  These findings contribute to a deeper understanding of the hadronic dynamics in \(\Lambda_b\) decays and provide valuable theoretical input that can be tested in ongoing and future experimental investigations, particularly in the search for exotic hadronic molecules involving open-charm mesons and strange baryons.

Overall, this study underscores the importance of considering both pseudoscalar and vector meson-baryon interactions to achieve a comprehensive picture of the heavy hadron spectrum and the mechanisms that give rise to exotic hadronic states near thresholds.

\section*{ACKNOWLEDGMENTS}
This work is partly supported by the National Natural Science
Foundation of China under Grants  No. 12405089 and No. 12247108 and
the China Postdoctoral Science Foundation under Grant
No. 2022M720360 and No. 2022M720359.  This work is also supported by the Spanish
Ministerio de Economia y Competitividad (MINECO) and European FEDER 
funds under Contracts No. FIS2017-84038-
C2-1-P B, PID2020- 112777GB-I00, and by Generalitat Valenciana under 
con- tract PROMETEO/2020/023. This project
has received funding from the European Union Horizon 2020 research and 
innovation programme under the program
H2020- INFRAIA-2018-1, grant agreement No. 824093 of the STRONG-2020 
project. This work is supported by the Spanish Ministerio de Ciencia e 
Innovacion (MICINN) under contracts PID2020-112777GB-I00, 
PID2023-147458NB- ´
C21 and CEX2023-001292-S; by Generalitat Valenciana under contracts 
PROMETEO/2020/023 and CIPROM/2023/59.
\bibliography{refs.bib} 
\end{document}